%% file: STAGE_Quantum_Photo_Booth_Paper.tex
\newcommand{\CLASSINPUTtoptextmargin}{0.75in}%
\newcommand{\CLASSINPUTbottomtextmargin}{1in}%
\newcommand{\CLASSINPUTinnersidemargin}{0.63in}%
\newcommand{\CLASSINPUToutersidemargin}{0.63in}%
\documentclass[conference,10pt,letterpaper]{IEEEtran}% 
\usepackage[hyphens]{url}
\usepackage{hyperref}
\hypersetup{breaklinks=true}

\usepackage{cite}
\usepackage{amsmath}% for double integral symbol in this template
\usepackage{graphicx}% for figures
\usepackage{multirow}% to allow multiple-row elements in tabular environment
\usepackage[none]{hyphenat}% turn off hyphenation to make text extraction and indexing easier
\usepackage{float}% better control of floating figures and tables
\usepackage{subfig}% for subfigures within figures
\usepackage{dblfloatfix}% fix to allow page-wide floats at bottom of page

\usepackage{t1enc}% allows access to various special characters
\usepackage{times}% change font to Nimbus Roman (based on Times Roman)
\input{IMS_modify_IEEEtran_18b_CTAN_V4c}

\begin{document}
%%%%%%%%%%%%%%%%%%%%%%%%%%%%%%%%%%%%%%%%%%%%%%%%%%%%%%%%%%%%%%%%%%%%%%%%%%%%%
% We use \raggedbottom to avoid latex adding vertical space around headings.
% This gives a better idea to the author about how much white space remains
% as the page limit is approached.
\raggedbottom
%
%%%%%%%%%%%%%%%%%%%%%%%%%%%%%%%%%%%%%%%%%%%%%%%%%%%%%%%%%%%%%%%%%%%%%%%%%%%%%
% PAPER TITLE AND AUTHOR BLOCK
%
% The paper title can use linebreaks \\ within to get better formatting if desired.
%
\title{Game Design Inspired by Quantum Physics: A Case Study on The Quantum Photo Booth}
%
% Next we define the author names and affiliations.
% Author names are listed using \IMSauthorblockNAME{} with comma separators between names.
% Affiliations are listed using \IMSauthorblockAFFIL{} with \\ separators between affiliations.
% Email addresses are listed using \IMSauthorblockEMAIL{} with comma separators between emails.
% See below for examples of each of these.
%
% Symbols marking author-affiliation relations are output using \IMSauthorrefmark{}.
%
% Next we typeset the authorblock either as visible text, or as an empty
% box of the same size, based on the value of the Blind Review Flag.
% Note that the Blind Review Flag also determines whether the Acknowledgements
% section is visible or invisible.
% To set the flag to Blind Review mode, simply uncomment the next line
\IMSthispaperforblindreview
% or to set the flag to Final Paper mode (with author block visible) then
% simply uncomment the next line:
\IMSthispaperforfinalpublication
%

%

%\IMSauthorblockNAME{Authors Name/s per 1st Affiliation (Author)}
%\IMSauthorblockA{line 1 (of Affiliation): dept. name of organization\\
%line 2: name of organization, acronyms acceptable\\
%line 3: City, Country\\
%line 4: Email: name@xyz.com}
%\author{\IEEEauthorblockN{Authors Name/s per 1st Affiliation (Author)}
%\IEEEauthorblockA{line 1 (of Affiliation): dept. name of organization\\
%line 2: name of organization, acronyms acceptable\\
%line 3: City, Country\\
%line 4: Email: name@xyz.com}
%\and
%\IEEEauthorblockN{Authors Name/s per 2nd Affiliation (Author)}
%\IEEEauthorblockA{line 1 (of Affiliation): dept. name of organization\\
%line 2: name of organization, acronyms acceptable\\
%line 3: City, Country\\
%line 4: Email: name@xyz.com}
%}

%\IMSauthor{%
%\IMSauthorblockNAME{% Author Names
%Sunanda Prabhu Gaunkar,\IMSauthorrefmark{\#}
%Denise Fischer,\IMSauthorrefmark{\#}
%Filip Rozp\k{e}dek,\IMSauthorrefmark{\#}
%Umang Bhatia,\IMSauthorrefmark{\#}
%\textsuperscript{\textsection}
%Shobhit Verma,\IMSauthorrefmark{\#}
%\textsuperscript{\textsection}
%Uri Zvi,\IMSauthorrefmark{\#}
%Nancy Kawalek\IMSauthorrefmark{\#}
%}% end of \IMSauthorblockNAME
%\\%

\IMSauthor{%
\IMSauthorblockNAME{% Author Names
Sunanda Prabhu Gaunkar,\IMSauthorrefmark{*}\textsuperscript{\textdagger}\textsuperscript{\textdaggerdbl}\textsuperscript{\textdollar}
Denise Fischer,\IMSauthorrefmark{*}\textsuperscript{\textdaggerdbl}\textsuperscript{\textdollar}
Filip Rozp\k{e}dek,\IMSauthorrefmark{*}\textsuperscript{\textdaggerdbl}
Umang Bhatia,\IMSauthorrefmark{*}\textsuperscript{\textsection}\textsuperscript{\textdaggerdbl}
Shobhit Verma,\IMSauthorrefmark{*}\textsuperscript{\textsection}\textsuperscript{\textdaggerdbl}
Ahit Kaan Tarhan,\IMSauthorrefmark{*}\textsuperscript{\textdollar}
Uri Zvi,\IMSauthorrefmark{*}\textsuperscript{\textdaggerdbl}
Nancy Kawalek\IMSauthorrefmark{*}\textsuperscript{\textdagger}\textsuperscript{\textdaggerdbl}\textsuperscript{\textdollar}
}% end of \IMSauthorblockNAME
\\%

\IMSauthorblockAFFIL{% Author Affiliations
\IMSauthorrefmark{*}STAGE Lab, Pritzker School of Molecular Engineering, University of Chicago, USA\\
%\IMSauthorrefmark{\$}Gollum World, New Zealand\\
%\IMSauthorrefmark{*}My Research, Australia\\
%\IMSauthorrefmark{\textasciicircum}SolarSolar, Australia
}% end of \IMSauthorblockAFFIL
%

%remove this\IMSauthorblockEMAIL{% Author Emails
%remove this\IMSauthorrefmark{1}spg@uchicago.edu,  \IMSauthorrefmark{2}uriz@uchicago.edu, %remove this\IMSauthorrefmark{3}dkfischer@uchicago.edu, \IMSauthorrefmark{4}sanskriti@uchicago.edu,  \IMSauthorrefmark{5}jmarkman@uchicago.edu, %remove this\IMSauthorrefmark{6}kawalek@uchicago.edu 
%remove this}% end of \IMSauthorblockEMAIL
%
}% end of \IMSauthor
%

% Next we make the title/author block using the information defined above.
\maketitle

\begingroup\renewcommand\thefootnote{\textdollar}
\footnotetext{These authors contributed to conceptual development of the STAGE Lab Quantum Casino.}
\endgroup

\begingroup\renewcommand\thefootnote{\textdaggerdbl}
\footnotetext{These authors contributed to development of The Quantum Photo Booth.}
\endgroup

\begingroup\renewcommand\thefootnote{\textsection}
\footnotetext{These authors contributed equally to development of The Quantum Photo Booth.}
\endgroup
\begingroup\renewcommand\thefootnote{\textdagger}
\footnotetext{To whom correspondence should be addressed. E-mail: spg@uchicago.edu; kawalek@uchicago.edu.}
\endgroup
%\begingroup\renewcommand\thefootnote{\textsection}
%\daggerfootnote{These authors contributed equally to this work.}
%\endgroup
%
%%%%%%%%%%%%%%%%%%%%%%%%%%%%%%%%%%%%%%%%%%%%%%%%%%%%%%%%%%%%%%%%%%%%%%%%%%%%%
% ABSTRACT paragraph.
% in the abstract paragraph.
%
\begin{abstract}

In this paper, we explain the conceptual development of the STAGE Lab Quantum Casino (a.k.a. the STAGE Lab Quantum Arcade), one of the Lab's most recent artistic endeavors about quantum physics. This work consists of a series of card and digital games and an interactive experience, exposing the public to quantum physics and minimizing learning barriers. Furthermore, we will also present a case study of the interactive experience, in the form of The Quantum Photo Booth. 

%The STAGE Lab is excited to build games about quantum physics because of several reasons. Quantum physics explains the counter-intuitive and surprising ways in which matter behaves at the subatomic level. It is an exciting and growing field that is devising solutions to society’s most intractable challenges, from economic growth to climate action, and health and well-being. However, the concepts of quantum physics might be too abstract and mathematical for the general public to learn. Furthermore, there exist limited resources, which aim to make this complex field accessible and understandable to the public. Our work addresses these problems by communicating the concepts of quantum physics in a way that is comprehensible and accessible to the general public.

The STAGE Lab Quantum Casino provides an entertaining and approachable experience for people of all ages to become familiar with quantum physics. By using core concepts of quantum physics as tools and strategies to overcome challenges that arise in gameplay, players gain an intuitive understanding of these concepts. These games provide players with a first-hand experience of the following quantum physics concepts: measurement, superposition, encryption, decoherence, and entanglement. Instead of teaching the concepts through a traditional classroom pedagogy, these games aim to invoke curiosity, spark moments of playfulness, and catalyze play-centric learning modalities. This paper provides a general overview of the development of the STAGE Lab Quantum Casino, focusing on The Quantum Photo Booth experience and how science is integrated into the very nature of the game development process in addition to its outcome.

\end{abstract}
%
%%%%%%%%%%%%%%%%%%%%%%%%%%%%%%%%%%%%%%%%%%%%%%%%%%%%%%%%%%%%%%%%%%%%%%%%%%%%%
% THE REST OF THE PAPER follows.
%
 \begin{figure*}
\centering
\includegraphics[width=175mm]{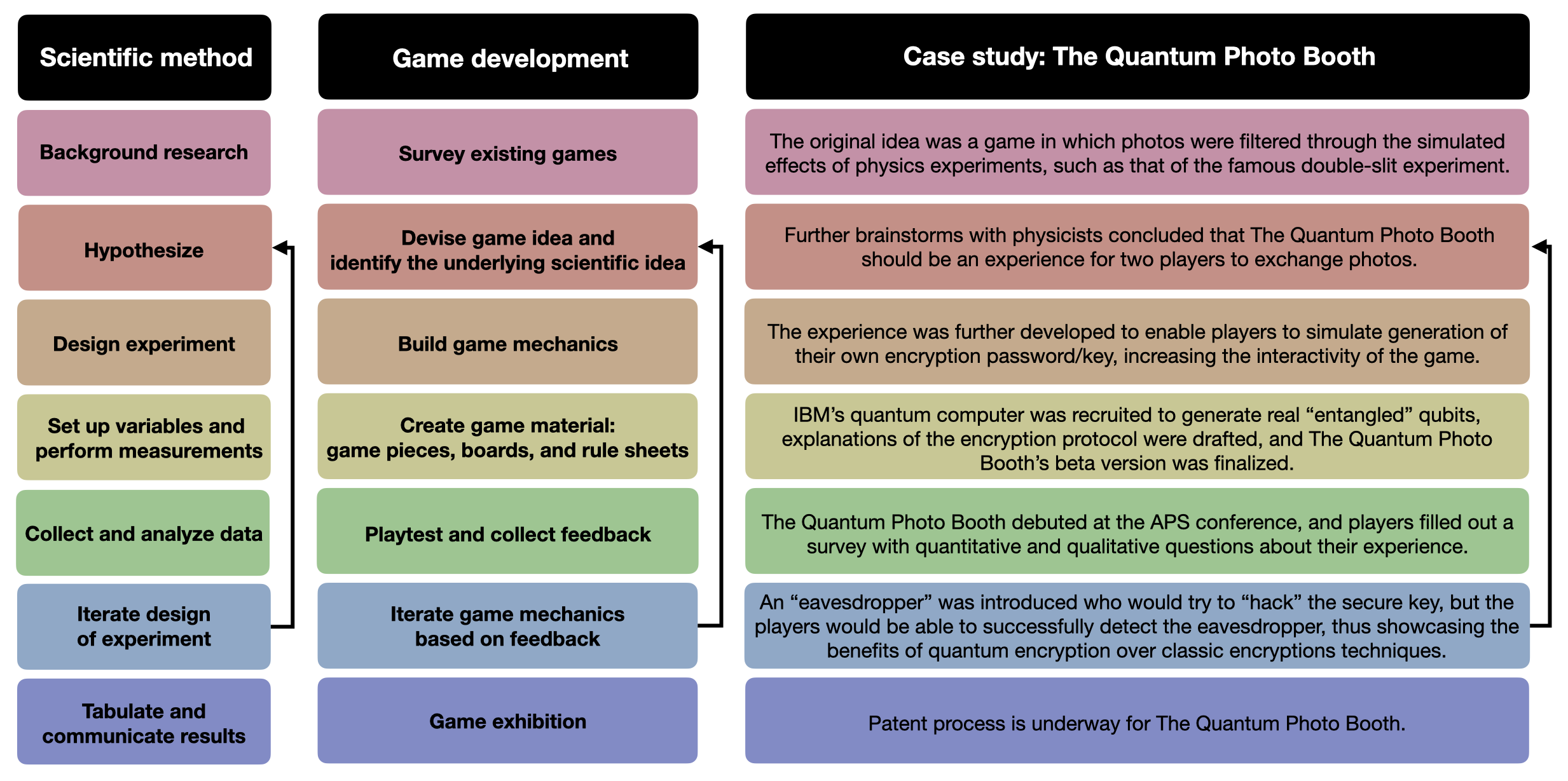}
\caption{The game development work process (Center block) with case study of The Quantum Photo Booth (Right block) and its parallels to the scientific method (Left block).}
\label{fig:figure1}
\end{figure*}
\section{Introduction to STAGE Lab}
STAGE — Scientists, Technologists and Artists Generating Exploration — is a full-scale laboratory embedded within the University of Chicago’s Pritzker School of Molecular Engineering. The STAGE Lab creates artistic works inspired by science and technology. 

The STAGE \cite{stage} Lab's distinct research focuses on creating and developing new artistic work inspired by science and technology. 
STAGE’s mission is to:
\begin{itemize}
\item Cultivate appreciation and collaboration between the two cultures of science and art;
\item Catalyze the development of art that depicts the technological age in which we live;
\item Promote an understanding of the sciences in the public arena;
\item Foster new and imaginative methods of storytelling;
\item Accomplish all of the above within an international community.
\end{itemize}

By cultivating collaboration between the arts and sciences, STAGE hopes to make science accessible to all, regardless of educational background. Typically, learning higher-level scientific concepts requires some foundation in science and mathematics, making them inaccessible to those who lack such a background. By integrating these concepts into an artistic form \cite{branchfromtree}, they can be learned, understood, and appreciated by all, thereby increasing the availability of scientific resources for public enjoyment and promoting scientific thinking. STAGE accomplishes this mission by utilizing the skills of a diverse group of lab members, including post-doctoral researchers, graduate students, and undergraduate students. Individuals at each level of education have a wide range of specialized knowledge in fields such as biology, chemistry, computer science, neuroscience, physics, molecular engineering, literature, philosophy, psychology, sociology, theatre, and visual arts. Both scientists and artists help create scientific content and craft narratives. Working closely with others from diverse areas of study and various countries exposes the creative team to numerous points of view and teaches them to communicate across disciplines and cultures. The work environment fosters openness and welcomes fresh ideas, leading to content without predeterminations based on personal biases. 
%Most importantly, experiencing different perspectives compels students to think and work in new ways, preparing them for greater current and future success. The work environment fosters openness and welcomes fresh ideas, leading to content without predeterminations based on personal biases. 
%Moreover, as a result of the diversity of lab members, the accessibility  of content for a broad audience is ensured.

STAGE focuses on portraying science through storytelling, which has a broad emotional appeal and intrinsic accessibility. By integrating science and storytelling, the impersonal and often obscure nature that many associate with science can be replaced by intrigue, wonder, and meaning. Children and adults can interact with concepts instead of merely reading about them. When conveyed through the language of art, science becomes less intimidating.

Science informs not only our subject matter but also our work process, which is inspired by the exploratory aspects of experimental science. Just as a scientific discovery emerges from experimentation, the narrative{\textemdash}whether in film, theatre, games, or other media{\textemdash}emerges from the process. The team iteratively creates "models" or prototypes, building the narrative step-by-step. The narrative unfolds, rich in nuance, surprises, and serendipitous breakthroughs. In this way, the work process heavily determines the outcome. To exemplify such a work process, this paper will detail the conceptual development of the STAGE Lab Quantum Casino with a specific case study of The Quantum Photo Booth.

\section{Origins of the STAGE Lab Quantum Casino}

The STAGE Lab Quantum Casino is one of STAGE’s most recent endeavors to illustrate quantum physics concepts through an interactive experience. The project began as an attempt to explain core concepts of quantum physics to non-scientists. This task proved challenging because the field explores abstract concepts invisible to the naked eye and relies on scientific jargon and advanced mathematics for explanations. In attempting to illuminate these concepts, STAGE quickly realized that an interactive and visual means would be helpful. This realization prompted the idea of a "quantum" casino, a physical space where audiences could play hand-held, digital, and interactive games inspired by and illustrative of quantum physics concepts. The casino space lends itself as a place where notions of randomness and uncertainty, which are fundamental concepts in casino games and quantum physics, could come together in an entertaining manner. The games sparked curiosity and offered exposure to the traditionally inaccessible concepts of quantum physics. More specifically, these games were designed to create engaging, interactive, and memorable experiences that afford a first-hand experience with the following core quantum physics topics: measurement, superposition, encryption, decoherence, and entanglement. The games elucidated these concepts because players utilized them as tools and strategies in playing the games.
%Even though the concepts of quantum physics are not observable in one’s day-to-day life, they explain the very nature of our universe.

In each game, the STAGE team worked to illustrate the scientific concepts through the game mechanics, forcing players to think like quantum physicists and  consider the same quantum physics concepts, obstacles, and goals that one would encounter in a laboratory. Utilizing these game mechanics is necessary for a player to understand, play, and win any of these games. Once the game was learned, so were the  concepts of quantum physics. Thus, with the desire to make cutting-edge research and science accessible to everyone, the design philosophy of the games aligned with that of STAGE. Moreover, the games aimed to invoke learning through player experience and interactivity instead of a science lecture. 

Creating new games to illustrate quantum physics was especially exciting because the discipline is so complex. Quantum physics, which attempts to explain the physical properties and behaviors of atomic and subatomic objects, has made astonishing progress in recent years with technological advances that can revolutionize computing, biomedical sensing, cryptography, and communication. However, learning quantum physics requires several prerequisites, such as a foundation in the sciences and mathematics, and is primarily taught at the undergraduate and higher levels. As a result, the concepts may be too elusive and abstract for the general public to learn. This difficulty is compounded by the limited availability of easily understandable resources and teaching materials that do not employ scientific jargon and equations. Additionally, more foundational quantum physics concepts are often left unexplained, leading to misconceptions and confusion. STAGE hopes to correct these misconceptions while also engaging new audiences with these concepts. 

Quantum physicists are developing cutting-edge technology for various applications, such as teleportation of information, unhackable communication, discovery of exotic materials, ultrafast computation, imaging of single-molecule dynamics, and life-changing biomedical tools. STAGE believes in sharing these ideas with a broad audience. Without significant efforts to engage and educate the public, a large fraction of society will remain disenfranchised and disconnected from this technological revolution, and this will in turn limit workforce development in this area. The STAGE Lab Quantum Casino was created to introduce the public to this complex new scientific field through games inspired by quantum physics.

 \section{Development of the STAGE Lab Quantum Casino}
 
 %The team began by playing and analyze existing games for potential to adapt for exploration of quantum core principles as well as invent, prototype and playtest adaptations and original game ideas. Following the STAGE Lab's unique work process, that mimics the exploratory aspects of experimental science research the team invented, prototyped and tested seven games based on core principles of quantum mechanics. These games include three card games: “Chicago Quant’em” (inspired by Texas Hold’em), “17” (inspired by Blackjack [21]), and “Quabble,” three digital games: “Qunnect 4,” “Tailspin,” and “Qubette” and one digital “experience”: “Quantum Photo Booth,” in collaboration with IBM. The team helped produce and run games playtests on campus and solicited and synthesized feedback to analyzed results from every playtest to improve upon the games.
 
 The development of the games followed the scientific method, using a top-down means to answer the question, "How can the concepts of quantum physics be clearly and accurately illustrated through games that are easy to understand for people of diverse ages and educational backgrounds?" The team began by playing and analyzing existing physics-based games for inspiration. This research of pre-existing games \cite{fuchs2020quantum,lin2020quantum,salimi2009investigation} was supplemented by brainstorming and inventing original game ideas. Upon devising a game idea, the team prototyped its rules, designed and playtested it with a  diverse group, and gathered feedback. The team created seven games:  
\begin{itemize}
\item Three card games: “Chicago Quant’em” (inspired by Texas Hold’em), “17” (inspired by Blackjack, also known as 21), and “Quabble.”
\item Three digital games: “Qunnect 4,” “Tailspin,” and “Qubette.”
\item One interactive “experience” created in collaboration with IBM: “The Quantum Photo Booth.”
\end{itemize}

After receiving feedback on the first version of each game from the playtesters, the team revised various aspects, including game mechanics, game design, and difficulty level. The team repeated this process of playtesting and revising numerous times, running the games with different groups of players and continuously incorporating the feedback to improve each version. The development of each game often involved the creation of different levels, to make the games engaging for people of various ages. Thus, the target audience for these games is anybody from age ten and up and from those with no scientific knowledge to scientific experts.

 \section{Findings/Results: Case Study of The Quantum Photo Booth}

One of the games the team developed is The Quantum Photo Booth, an interactive experience created in collaboration with IBM. Following the scientific method, development went through multiple cycles of hypothesis, experimentation, and data analysis (Fig.~\ref{fig:figure1}) with the goal of sending images between two players in a fun and interactive manner while learning some quantum physics. The first idea was to filter photos through the simulated effects of physics experiments, such as that of the famous quantum physics double-slit experiment. However, the team could not find a way to implement this experiment in a way that would elucidate the quantum mechanical interference effect.
%After taking photos in The Quantum Photo Booth, quantum bits (the fundamental particle in quantum physics, also known as qubits) generated by IBM’s quantum computer would interfere with each other and distort the pixels, thereby illustrating the quantum physics principle of wave-particle duality. However, the team quickly realized the interfering qubits would not meaningfully distort the photo, preventing the user from noticing a significant difference between the "quantum" photo and a standard photo. As a result, the quantum physics aspect of the experience would be unclear, leaving players without a clear understanding of wave-particle duality.

While brainstorming with a lab that conducts quantum computing research \cite{david, liang}, the team then transformed The Quantum Photo Booth from an experience that illustrates wave-particle duality to one that illustrates "Quantum Key Distribution (QKD) \cite{bennett2014quantum}." This version used IBM's quantum computer to create a key used for decryption. In our visualization, this key looked like a "white noise" image \cite{white}. The experience began with players in The Quantum Photo Booth being presented with encrypted images. Then, the white noise image was placed on top of the encrypted image to reveal the original photo. During implementation, however, the team struggled to determine how this key would be presented. The large file size that was created in generating a truly random key presented an obstacle. However, if compressed, the key wouldn’t work as intended. Despite the difficulties, the team continued to pursue this idea of quantum encryption, finding other ways to incorporate it into The Quantum Photo Booth. Ultimately, the team shifted focus to the difference between quantum encryption and classical encryption, namely how quantum encryption is much more secure and unhackable than classical encryption. The visualization of this distinction became the new educational goal of The Quantum Photo Booth. 

%In this experience,  
\begin{figure}
\centering
\includegraphics[width=85mm]{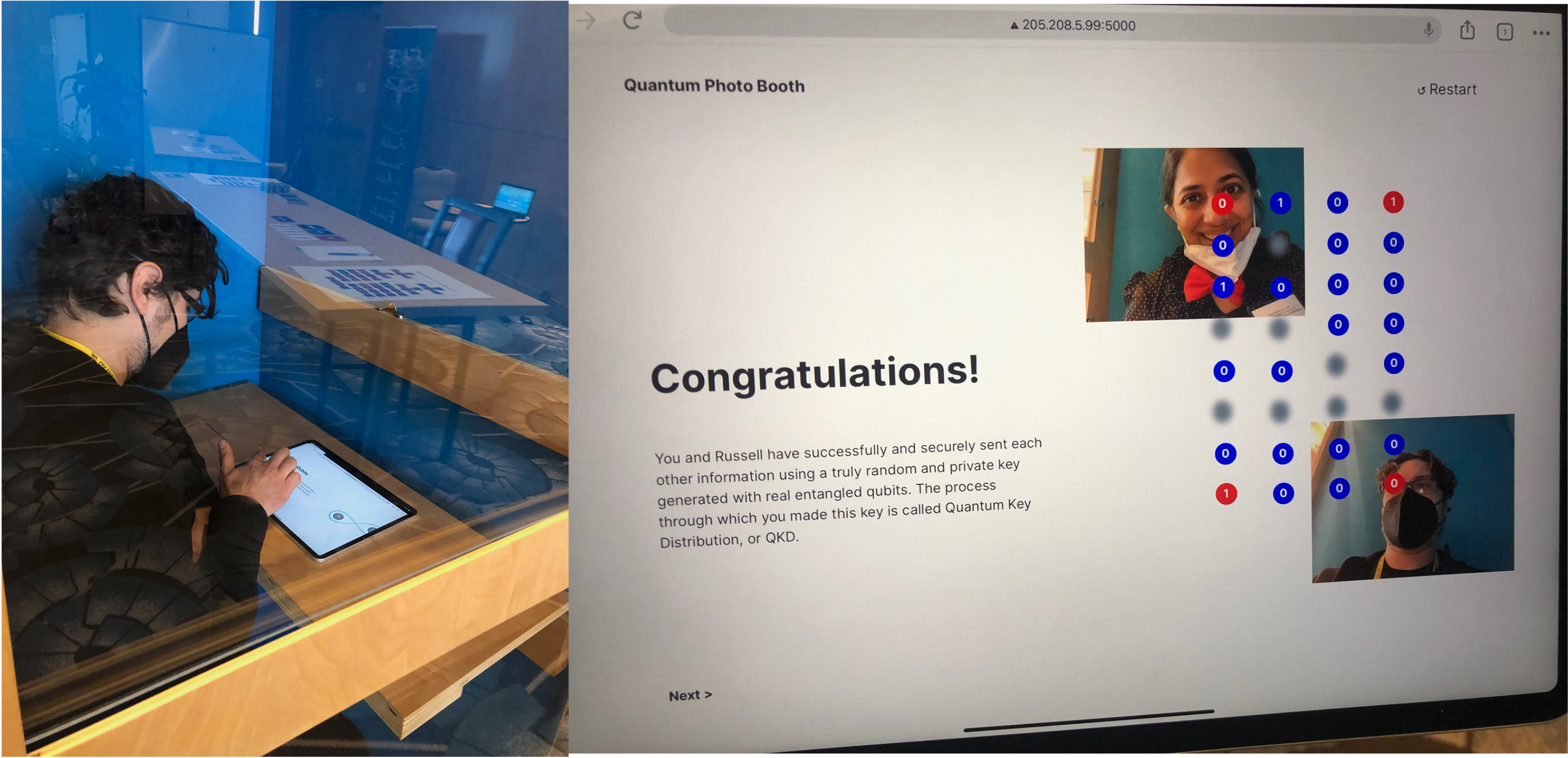}
\caption{(Image at left) The STAGE Lab Quantum Photo Booth demonstration at the 2022 American Physical Society March meeting in Chicago, IL, showing a participant using  the interactive panel inside The Quantum Photo Booth. (Image at right) Final screen displaying the successful viewing of photos between the two players using  the key they generated with the help of IBM’s entangled qubits.}
\label{fig:aps}
\end{figure}

In its final form, The Quantum Photo Booth (Fig.~\ref{fig:aps}) became an interactive experience demonstrating the principles of QKD. Two users play simultaneously and send a photo to each other. To send the photos securely, the players use quantum entanglement—a unique property of quantum physics—to create the most secure encryption key (an unhackable password). Specifically, the players send each other a photo and simulate encrypting it using a quantum encryption key created by measuring IBM-generated “entangled” qubits (made with the help of IBM’s quantum computer). During the experience, the players were guided through the steps of key generation, encouraging them to interact with the technology and gain a better understanding of the principles of QKD. While generating the key, players learn about quantum behavior, such as superposition and entanglement, concepts explained through this experience of developing this encryption key. To complete the experience, players view their partner's photo with this key. Initial feedback from the playtesting of The Quantum Photo Booth in its final version was positive overall, with players describing it as exciting, informative, and engaging. Further user data  across various demographics will be collected in future playtesting to gauge the success of the Quantum Photo Booth in communicating QKD. With scaling for distribution, The Quantum Photo Booth aims to generate widespread public interest in quantum technologies and inspire the next generation of the quantum workforce.

\section{Future Steps}

Currently, STAGE is working on incorporating feedback (Exemplified in the Appendix) received from playtests for The Quantum Photo Booth. Most recently, the team presented The Quantum Photo Booth at the 2022 American Physical Society March meeting [9] and received feedback from the attendees. After incorporating the feedback, the team continues testing this experience with other audiences for additional suggestions and input. Once all feedback is included and all the games, including The Quantum Photo Booth, are finalized, the team hopes to distribute them across the country, allowing people of all ages and interests to learn quantum physics from their homes. The team will continue creating new artistic endeavors using technology to communicate complex scientific ideas to the public and foster appreciation towards the importance and beauty of science.

\section*{Acknowledgments}
%In the future, the team plans on continuing our work on not only our quantum games and theatre production, but also by finding additional art medias helpful for communicating science. Currently, we are working on incorporating feedback into our quantum games that we received at our recent presentation at the American Physical Society conference in March 2022. After incorporation, we plan on testing the games with additional audiences for continued feedback. After completion of the quantum games, we hope to distribute them nationally, thereby allowing people of all ages and interests to learn quantum physics from their own homes. Regarding our theater production, we plan to continue developing our storyline through improvisational practices. Eventually, we plan to present the production in front of a live audience regionally, thereby sharing quantum physics in a new, heartwarming manner. Some additional goals for both of these projects and beyond include the increased incorporation of technology in our art.
% the following macro \IMSdisplayacksection makes its argument visible for the
% final paper, but hides its argument for the blind-review paper while keeping
% a blank vertical space of the same size.  The actual text of the acknowledgement
% is defined inside the macro \IMSacktext so that multiple paragraphs will be
% displayed or hidden correctly.
%\newcommand{\IMSacktext}
{

The STAGE Lab Quantum Casino is supported by the National Science Foundation grant NSF DMR-1830704.
The authors would like to acknowledge The Quantum Photo Booth team comprising front- and back-end developer and UI/UX advisor Dustin Heffron, UI/UX developer Michelle Klonsinski, additional consultants for the Quantum Photo Booth Sanskriti Chitransh and Rhys Povey, science advisors Dr. Liang Jiang and Dr. David Awschalom, and UChicago project management team Craig Hamill, Mary Pat McCullough, and Roell Schmidt. The authors thank IBM, and Iskandar Sitdikov and Olivia Lanes of IBM, for providing entangled qubits. The authors also thank the entire STAGE Lab game design team for developing this Quantum Casino.

}
%\vfill\null

%\IMSdisplayacksection{\IMSacktext}% end of \IMSdisplayacksection

%\begin{figure}
%\centering
%\includegraphics[width=85mm]{figures/aps.png}
%\caption{(Image at left) STAGE lab Quantum Photo Booth demonstration at the 2022 American Physical Society March meeting in Chicago, IL, showing a participant using  the interactive panel inside the Photo Booth. (Image at right) Final screen displaying the successful decryptions of photos between the two players using  the key they generated with the help of IBM’s entangled qubits.}
%\label{fig:aps}
%\end{figure}

%%%%%%%%%%%%%%%%%%%%%%%%%%%%%%%%%%%%%%%%%%%%%%%%%%%%%%%%%%%%%%%%%%%%%%%%%%%%%

%\bibliography{IEEEabrv,IEEEexample}
%\begin{thebibliography}{1}
%  \bibitem{stage}
%  https://stage.pme.uchicago.edu
%  \bibitem{}
%  The Integration of the Humanities and Arts with Sciences, Engineering, and Medicine in Higher Education
%  \bibitem{qiskit}
% https://qiskit.org/textbook/ch-algorithms/quantum-key-distribution.html
% \bibitem{pme}
% https://pme.uchicago.edu/news/betting-quantum-how-one-pme-lab-uses-game-design-explain-physics-most-complex-laws
% 
%
%\end{thebibliography}
\bibliographystyle{IEEEtran}
\bibliography{Sunandareferences}

\appendix
\label{appendix}
\section {Appendix Participant feedback about The Quantum Photo Booth}
The following feedback was provided by players testing iterations of The Quantum Photo Booth:
\begin{itemize}
\item “The IBM Photo Booth showed me how quantum physics can improve encryption. Did I actually get qubits right there?” (Player demographic: Student studying science, not quantum scientist; age range 26 – 41)
\item “I think The Photo Booth’s text was very informative in an audience-friendly way.” (Player demographic: General public, age range 58 – 76)
\item “I felt more excited about the ways we can engage the public concerning quantum science. The implications of quick QKD became more visceral after playing The Quantum Photo Booth. It makes me want to delve deeper into finding practical molecular qubits, and also makes me wonder as to future applications outside the ones often discussed.” (Player demographic: Student/teacher quantum scientist, age range 26 – 41)
\item “I’m much more interested in learning about quantum physics after playing the games. It has me thinking about potential social media content for my quantum start-up. I liked the interactive experience of The Quantum Photo Booth.” (Player demographic: Student artist,  age range 10 – 25)
\end{itemize}

\end{document}

%% file: IMS_modify_IEEEtran_18b_CTAN_V4c.tex
\makeatletter

\def\@IMSauthorblockNAMEstyle{\normalfont\IMSauthorsize}
\def\@IMSauthorblockAFFILstyle{\normalfont\IMSaffilsize}
\def\@IMSauthorblockEMAILstyle{\normalfont\IMSaffilsize}
\def\IMSauthorblockNAME#1{%
\relax\@IMSauthorblockNAMEstyle%
#1%
}%
\def\IMSauthorblockAFFIL#1{%
\relax\@IMSauthorblockAFFILstyle%
\vskip\@IEEEauthorblockAtopspace
#1%
}%
\def\IMSauthorblockEMAIL#1{%
\relax\@IMSauthorblockEMAILstyle%
\vskip\@IEEEauthorblockAtopspace
#1%
}%
\newcommand{\IMSauthor}[1]{%
\ifIsBlindReviewVersion%
\author{\phantom{\parbox{\textwidth}{\center\relax#1}}}%
\else%
\author{\parbox{\textwidth}{\center\relax#1}}%
\fi%
}%
\newif\ifIsBlindReviewVersion
\def\IMSthispaperforblindreview{\IsBlindReviewVersiontrue}
\def\IMSthispaperforfinalpublication{\IsBlindReviewVersionfalse}
\def\@maketitle{\newpage
\bgroup\par\addvspace{0.5\baselineskip}\centering%
\ifCLASSOPTIONtechnote% technotes
   {\bfseries\large\@IEEEcompsoconly{\sffamily}\@title\par}\vskip 1.3em{\lineskip .5em\@IEEEcompsoconly{\sffamily}\@author
   \@IEEEspecialpapernotice\par{\@IEEEcompsoconly{\vskip 1.5em\relax
   \@IEEEtitleabstractindextextbox{\@IEEEtitleabstractindextext}\par
   \hfill\@IEEEcompsocdiamondline\hfill\hbox{}\par}}}\relax
\else% not a technote
   \vskip0.2em{\IMStitlesize\ifCLASSOPTIONtransmag\bfseries\LARGE\fi\@IEEEcompsoconly{\sffamily}\@IEEEcompsocconfonly{\normalfont\normalsize\vskip 2\@IEEEnormalsizeunitybaselineskip
   \bfseries\Large}\@title\par}\vskip1.0em\par% CAUSAL PRODUCTIONS change on this line
   \ifCLASSOPTIONconference%
      {\@IEEEspecialpapernotice\mbox{}\vskip\@IEEEauthorblockconfadjspace%
       \mbox{}\hfill\begin{@IEEEauthorhalign}\@author\end{@IEEEauthorhalign}\hfill\mbox{}\par}\relax
   \else% peerreviewca, peerreview or journal
      \ifCLASSOPTIONpeerreviewca
         {\@IEEEcompsoconly{\sffamily}\@IEEEspecialpapernotice\mbox{}\vskip\@IEEEauthorblockconfadjspace%
          \mbox{}\hfill\begin{@IEEEauthorhalign}\@author\end{@IEEEauthorhalign}\hfill\mbox{}\par
          {\@IEEEcompsoconly{\vskip 1.5em\relax
           \@IEEEtitleabstractindextextbox{\@IEEEtitleabstractindextext}\par\hfill
           \@IEEEcompsocdiamondline\hfill\hbox{}\par}}}\relax
      \else% journal, peerreview or transmag
         \ifCLASSOPTIONtransmag
           {\@IEEEspecialpapernotice\mbox{}\vskip\@IEEEauthorblockconfadjspace%
            \mbox{}\hfill\begin{@IEEEauthorhalign}\@author\end{@IEEEauthorhalign}\hfill\mbox{}\par
           {\vspace{0.5\baselineskip}\relax\@IEEEtitleabstractindextextbox{\@IEEEtitleabstractindextext}\vspace{-1\baselineskip}\par}}\relax
         \else% journal or peerreview
           {\lineskip.5em\@IEEEcompsoconly{\sffamily}\sublargesize\@author\@IEEEspecialpapernotice\par
           {\@IEEEcompsoconly{\vskip 1.5em\relax
            \@IEEEtitleabstractindextextbox{\@IEEEtitleabstractindextext}\par\hfill
            \@IEEEcompsocdiamondline\hfill\hbox{}\par}}}\relax
         \fi
      \fi
   \fi
\fi\par\addvspace{0.0\baselineskip}\egroup}% CAUSAL PRODUCTIONS change on this line, reduce the vspace from 0.5\baselineskip to 0.0

\def\IMStitlesize{\@setfontsize{\IMStitlesize}{18}{21pt}}% CAUSAL PRODUCTIONS change on this line
\def\IMSauthorsize{\@setfontsize{\IMSauthorsize}{12}{13pt}}% CAUSAL PRODUCTIONS change on this line
\def\IMSaffilsize{\@setfontsize{\IMSaffilsize}{12}{13pt}}% CAUSAL PRODUCTIONS change on this line
\def\IMScaptionsize{\@setfontsize{\IMScaptionsize}{8}{9pt}}% CAUSAL PRODUCTIONS change on this line
\def\IMSbibsize{\@setfontsize{\IMSbibsize}{8}{9pt}}% CAUSAL PRODUCTIONS change on this line

\def\@IEEEauthorblockNstyle{\IMSauthorsize\@IEEEcompsocnotconfonly{\sffamily}\@IEEEcompsocconfonly{\large}}%CAUSAL PRODUCTIONS removed sublargesize to get correct IMSauthorsize
\def\@IEEEauthorblockAstyle{\IMSaffilsize\@IEEEcompsocnotconfonly{\sffamily}\@IEEEcompsocconfonly{\itshape}\@IEEEcompsocconfonly{\large}}%CAUSAL PRODUCTIONS removed normalsize to get correct IMSaffilsize
\def\@IEEEauthordefaulttextstyle{\IMSauthorsize\@IEEEcompsocnotconfonly{\sffamily}\sublargesize}%CAUSAL PRODUCTIONS

\def\thebibliography#1{\section*{\refname}%
    \addcontentsline{toc}{section}{\refname}%
    \IMSbibsize\@IEEEcompsocconfonly{\small}\vskip 0.3\baselineskip plus 0.1\baselineskip minus 0.1\baselineskip% CAUSAL PRODUCTIONS change on this line
    \list{\@biblabel{\@arabic\c@enumiv}}%
    {\settowidth\labelwidth{\@biblabel{#1}}%
    \leftmargin\labelwidth
    \advance\leftmargin\labelsep\relax
    \itemsep \IEEEbibitemsep\relax
    \usecounter{enumiv}%
    \let\p@enumiv\@empty
    \renewcommand\theenumiv{\@arabic\c@enumiv}}%
    \let\@IEEElatexbibitem\bibitem%
    \def\bibitem{\@IEEEbibitemprefix\@IEEElatexbibitem}%
\def\newblock{\hskip .11em plus .33em minus .07em}%
\ifCLASSOPTIONtechnote\sloppy\clubpenalty4000\widowpenalty4000\interlinepenalty100%
\else\sloppy\clubpenalty4000\widowpenalty4000\interlinepenalty500\fi%
    \sfcode`\.=1000\relax}

\long\def\@makecaption#1#2{%
\ifx\@captype\@IEEEtablestring%
\par\@IEEEtabletopskipstrut% strut used to align table caption with facing column
\else
\@IEEEfigurecaptionsepspace
\fi
\setbox\@tempboxa\hbox{\normalfont\IMScaptionsize {#1.}\nobreakspace\nobreakspace #2}%
\ifdim \wd\@tempboxa >\hsize%
\setbox\@tempboxa\hbox{\normalfont\IMScaptionsize {#1.}\nobreakspace\nobreakspace}%
\parbox[t]{\hsize}{\normalfont\IMScaptionsize\noindent\unhbox\@tempboxa#2}%
\else
\ifCLASSOPTIONconference \hbox to\hsize{\normalfont\IMScaptionsize\hfil\box\@tempboxa\hfil}%
\else \hbox to\hsize{\normalfont\IMScaptionsize\box\@tempboxa\hfil}%
\fi\fi
\ifx\@captype\@IEEEtablestring%
\@IEEEtablecaptionsepspace
\else
\fi}

\newlength\tablecaptiontotableskip
\newlength\figuretocaptionskip
\def\@IEEEfigurecaptionsepspace{\vskip\figuretocaptionskip\relax}%
\def\@IEEEtablecaptionsepspace{\vskip\tablecaptiontotableskip\relax}%

\def\abstract{\normalfont%
\@IEEEabskeysecsize\bfseries\textit{\abstractname}\,\bfseries\textit{---}\,%
\@IEEEgobbleleadPARNLSP}%

\def\IEEEkeywords{\normalfont%
\@IEEEabskeysecsize\bfseries\textit{\IEEEkeywordsname}\,\bfseries\textit{---}\,%
\@IEEEgobbleleadPARNLSP}%
\def\endIEEEkeywords{\relax\vspace{0.67ex}%
\par\if@twocolumn\else\endquotation\fi%
\normalsize\normalfont}%

\DeclareRobustCommand*{\IMSauthorrefmark}[1]{\raisebox{0pt}[0pt][0pt]{\textsuperscript{\footnotesize{#1}}}}%
\def\@IEEEauthorblockNtopspace{0ex}
\def\@IEEEauthorblockAtopspace{1mm}
\def\tablename{Table}% IMS: was TABLE in IEEETran, but we are now using capitalized caption text not smallcaps
\def\thetable{\arabic{table}}% IMS: Tables are numbered using arabic not roman
\def\IEEEkeywordsname{Keywords}% use Keywords instead of Index Terms
\def\subsubsection{\@startsection{subsubsection}{3}{\z@}{1.5ex plus 1.5ex minus 0.5ex}%
{0.7ex plus .5ex minus 0ex}{\normalfont\normalsize\itshape}}%
\def\@seccntformat#1{\csname the#1dis\endcsname\relax}% moved the spacer \hskip 0.5em to individual handlers below
\def\thesectiondis{\thesection.\hskip 0.5em}%					I.	% CAUSAL PRODUCTIONS: moved the hskip spacer from \@seccntformat to here
\def\thesubsectiondis{{\hbox to\parindent{\Alph{subsection}.}}}%		B.	% CAUSAL PRODUCTIONS: indent the subsection name to match paragraph indent
\def\thesubsubsectiondis{{\hbox to \parindent{\arabic{subsubsection})}}}%	3)	% CAUSAL PRODUCTIONS: indent the subsubsection name to match paragraph indent
\def\theparagraphdis{{\hbox to \parindent{\alph{paragraph})}}}%			d)	% CAUSAL PRODUCTIONS: indent the subsubsubsection name to match paragraph indent
\IEEEilabelindentA \parindent
\IEEEilabelindent \IEEEilabelindentA
\IEEEelabelindent \parindent
\IEEEdlabelindent \parindent
\IEEElabelindent \parindent
\newlength\@IMSparindent
\newcommand\IMSdisplayacksection[1]{%
\ifIsBlindReviewVersion%
\noindent\phantom{\parbox[t]{\columnwidth}{\normalbaselines\setlength{\parindent}{\@IMSparindent}{#1}\strut}}%\IMSacktext
\else%
\noindent\parbox[t]{\columnwidth}{\normalbaselines\setlength{\parindent}{\@IMSparindent}{#1}\strut}%
\fi%
}%

\makeatother

%% file: STAGE_Quantum_Photo_Booth_Paper.bbl
\begin{thebibliography}{1}
\providecommand{\url}[1]{#1}
\csname url@samestyle\endcsname
\providecommand{\newblock}{\relax}
\providecommand{\bibinfo}[2]{#2}
\providecommand{\BIBentrySTDinterwordspacing}{\spaceskip=0pt\relax}
\providecommand{\BIBentryALTinterwordstretchfactor}{4}
\providecommand{\BIBentryALTinterwordspacing}{\spaceskip=\fontdimen2\font plus
\BIBentryALTinterwordstretchfactor\fontdimen3\font minus
  \fontdimen4\font\relax}
\providecommand{\BIBforeignlanguage}[2]{{%
\expandafter\ifx\csname l@#1\endcsname\relax
\typeout{** WARNING: IEEEtran.bst: No hyphenation pattern has been}%
\typeout{** loaded for the language `#1'. Using the pattern for}%
\typeout{** the default language instead.}%
\else
\language=\csname l@#1\endcsname
\fi
#2}}
\providecommand{\BIBdecl}{\relax}
\BIBdecl

\bibitem{stage}
``{Scientists, Technologists and Artists Generating Exploration (STAGE) Lab},''
  \url{https://stage.pme.uchicago.edu}.

\bibitem{branchfromtree}
\BIBentryALTinterwordspacing
D.~Skorton, ``Branches from the same tree: The case for integration in higher
  education,'' \emph{Proceedings of the National Academy of Sciences}, vol.
  116, no.~6, pp. 1865--1869, 2019. [Online]. Available:
  \url{https://www.pnas.org/doi/abs/10.1073/pnas.1807201115}
\BIBentrySTDinterwordspacing

\bibitem{fuchs2020quantum}
F.~G. Fuchs, V.~Falch, and C.~Johnsen, ``Quantum poker---a game for quantum
  computers suitable for benchmarking error mitigation techniques on nisq
  devices,'' \emph{The European Physical Journal Plus}, vol. 135, no.~4, p.
  353, 2020.

\bibitem{lin2020quantum}
J.~X. Lin, J.~A. Formaggio, A.~W. Harrow, and A.~V. Natarajan, ``Quantum
  blackjack: Advantages offered by quantum strategies in communication-limited
  games,'' \emph{Physical Review A}, vol. 102, no.~1, p. 012425, 2020.

\bibitem{salimi2009investigation}
S.~Salimi and M.~Soltanzadeh, ``Investigation of quantum roulette,''
  \emph{International Journal of Quantum Information}, vol.~7, no.~03, pp.
  615--626, 2009.

\bibitem{david}
``{Pritzker School of Molecular Engineering Quantum Electronics Lab,
  UChicago},'' \url{https://pme.uchicago.edu/group/awschalom-group}.

\bibitem{liang}
``{Pritzker School of Molecular Engineering Quantum Computation Lab,
  UChicago},'' \url{https://pme.uchicago.edu/group/jiang-group}.

\bibitem{bennett2014quantum}
C.~H. Bennett and G.~Brassard, ``Quantum cryptography: Public key distribution
  and coin tossing,'' \emph{Theoretical computer science}, vol. 560, pp. 7--11,
  2014.

\bibitem{white}
``{White noise image},'' \url{https://en.wikipedia.org/wiki/White_noise}.

\end{thebibliography}
